\documentclass[aps, prd, showpacs, superscriptaddress, 
nofootinbib, twocolumn,floatfix,preprintnumbers
]{revtex4}

\usepackage{comment}
\usepackage{graphicx}\usepackage{amsmath}\usepackage{natbib}
\usepackage{times}
\usepackage{amsmath}
\usepackage{CJK}
 
\def\beq{\begin{equation}} 
\def\eeq{\end{equation}} 
\def\beqar{\begin{eqnarray}} 
\def\eeqar{\end{eqnarray}}

\def\msol{M_\odot}

\def\csnr{{\cal R}_{\rm tot}} 
\def\vis{{\cal R}_{\rm vis}}
\def\invis{{\cal R}_{\rm invis}}

\def\enu{\epsilon}

\def\etot{{\cal E}_{\nu,\rm tot}}

\def\mag#1{#1^{\rm mag}}

\def\mlim{m_{\rm lim}^{\rm sn}} 
\def\oscan{\Delta \Omega_{\rm scan}}

\def\nuebar{\bar{\nu}_e}

\def\la{\mathrel{\mathpalette\fun <}}
\def\ga{\mathrel{\mathpalette\fun >}}
\def\fun#1#2{\lower3.6pt\vbox{\baselineskip0pt\lineskip.9pt
  \ialign{$\mathsurround=0pt#1\hfil##\hfil$\crcr#2\crcr\sim\crcr}}}

\begin{document} 
\begin{CJK*}{Bg5}{bsmi}

\title{Synoptic Sky Surveys and the Diffuse Supernova Neutrino Background: \\
 Removing Astrophysical Uncertainties and Revealing Invisible Supernovae} 

\author{Amy Lien (³s¶®µY)}
\affiliation{Department of Astronomy, University of Illinois, Urbana, 
  Illinois 61801, USA} 
\author{Brian D. Fields} 
\affiliation{Department of Astronomy, University of Illinois, Urbana, 
  Illinois 61801, USA} 
\affiliation{Department of Physics, University of Illinois, Urbana, 
  Illinois 61801, USA} 
\author{John F. Beacom}
\affiliation{Center for Cosmology and Astro-Particle Physics, Ohio State University, Columbus, 
  Ohio 43210, USA}
\affiliation{Department of Physics, Ohio State University, Columbus,
   Ohio 43210, USA}
\affiliation{Department of Astronomy, Ohio State University, Columbus, 
  Ohio 43210, USA}
 
\begin{abstract} 
The cumulative (anti)neutrino production from
all core-collapse supernovae within our cosmic horizon
gives rise to the diffuse supernova neutrino background (DSNB),
which is on the verge of detectability.
The observed flux depends on supernova physics, but also
on the cosmic history of supernova explosions;
currently, the cosmic supernova rate introduces a substantial
($\pm 40\%$)
uncertainty, largely through its absolute normalization.
However, a new class of wide-field, repeated-scan (synoptic)
optical sky surveys
is coming online, and will map the sky in the time domain
with unprecedented depth, completeness, and dynamic range.
We show that these surveys will obtain the cosmic supernova
rate by {\em direct counting},
in an unbiased way and with high statistics,
and thus will allow for precise 
predictions of the DSNB.  
Upcoming sky surveys will substantially
reduce the uncertainties in the
DSNB source history to an anticipated $\pm 5\%$
that is dominated by systematics,
so that the observed high-energy flux
thus will test supernova neutrino physics.
The portion of the universe ($z \la 1$) 
accessible to upcoming sky surveys includes 
the progenitors of a large fraction 
($\simeq 87\%$)
of the expected 10 -- 26 MeV DSNB event rate.
We show that 
precision determination of the (optically detected) cosmic supernova history
will also make the DSNB into a strong probe of an extra flux of neutrinos from
optically {\em invisible}
supernovae, which may be unseen either due to unexpected large
dust obscuration in host galaxies, or because some core-collapse
events proceed directly to black hole formation and fail to
give an optical outburst.
\end{abstract} 


\maketitle
\end{CJK*}
 
\section{Introduction} 

\label{sect:intro}

Core-collapse supernovae are the
spectacular outcome of the violent deaths of massive stars. 
These events, which
include 
Type II, Type Ib, and Type Ic supernovae, 
are in a real sense ``neutrino bombs''
in which the production and emission of 
neutrinos dominates the
dynamics and energetics.
This basic picture now rests on firm observational footing
in light of the detection 
of neutrinos from SN 1987A \cite{hirata,bionta}.
Thus all massive star deaths -- certainly those that yield
optical explosions, and
even ``invisible'' events that do not -- are powerful
neutrino sources.  Yet only the very closest events
can be individually detected by neutrino observatories,
leading to burst rates so small that no new events have
been seen in more than two decades.

All core-collapse events within the observable
universe emit neutrinos whose ensemble constitutes
the diffuse supernova neutrino background (DSNB)
\footnote{
Type Ia supernovae do not 
have substantial neutrino emission $> 10$ MeV,
but an intriguing alternative fate of accreting
white dwarfs is the accretion-induced collapse (AIC)
to a neutron star.
Ref.~\cite{fryer09} 
suggests 
the AIC events can also produce neutrino emission
similar to core-collapse events.
If so, AIC events would contribute to the DSNB
and to optically visible outbursts.
However, these AIC events have not yet been observationally confirmed
and the expected AIC rate is much lower than that of 
core-collapse events. Therefore AIC neutrinos should not 
greatly change our results.
}
\cite{group1,group2,group3,daigne05,Strigari04,yuksel05,lunardini06,horiuchi}.
Core-collapse supernovae produce all three active neutrino
species (and their antineutrinos), all in roughly equal numbers.
However, for the foreseeable future only the $\bar{\nu}_e$
flux can be detected above backgrounds present on Earth.
Specifically, the DSNB dominates the (anti)neutrino
flux at Earth in the
$\sim 10-26$ MeV energy range,
and has long been a tantalizing
signal that has become a
topic of intense interest 
(e.g., \cite{group1,group2,group3,daigne05,Strigari04,yuksel05,lunardini06,horiuchi}).
Until now no DSNB signal
has been detected, which
set an upper bound on the DSNB flux.
Super-Kamiokande (Super-K) set 
the upper limit to be 
$1.2 \, \rm cm^{-2} \, s^{-1}$ above 19.3 MeV of the neutrino energy
\cite{malek}. 
However, this limit is already close to 
theoretical prediction and thus
Super-K is expecting to
detect the first DSNB signal within the
next several years.

Recently,
Ref.~\cite{horiuchi} considered a variety of 
complementary indicators of the cosmic supernova rate, 
and concluded that the DSNB is no more than a factor $\sim 2-4$
below the 2003 Super-K limit \cite{malek}. Moreover these authors point out
that if Super-K is enhanced with gadolinium 
to tag detector background events
\cite{beacom03}, the 
resulting enhanced sensitivity at 10 -- 18 MeV
should lead to a firm DSNB detection.

In light of the impending DSNB detection it is 
imperative to quantify the uncertainties in the prediction
and to reduce these as much as possible.
The predicted flux depends crucially on:
(a) supernova neutrino physics, via
the emission per supernova; and
(b) the cosmic history of core-collapse supernovae, via
the cosmic supernova rate (hereafter, CSNR).
Our emphasis in the present paper is on the CSNR,
which has begun to be measured 
in a qualitatively new way by 
``synoptic'' surveys.
These new campaigns repeatedly scan the sky
with a certain fields of view
and high sensitivity.
Pioneering synoptic surveys 
are already in hand and have shown the power of this technique.
To date, these surveys have reported the detection of 
several hundreds of
supernovae in total, including both Type Ia and core-collapse events
\cite{essence,sdss-sn, snls, hst, TSS, pq, ptf, catalina}.
Future surveys, such as DES, Pan-STARRS, and LSST,
should find
$> 10^4\ \rm CCSNe \ yr^{-1}$,
eventually with detection rates of
$> 10^5\ \rm CCSNe \ yr^{-1}$ 
based on their depths and large fields of view \cite{lf}.
Current predictions show that 
these data will provide an absolute measurement of
the CSNR to high statistical precision out to $z \sim 1$
\cite{lf,LSSTsb}.
Note that observations
seem to suggest that Type IIn supernovae 
are intrinsically the most luminous core-collapse type
\cite{richardson}, and
therefore would contribute 
to most of the detections at $z \gtrsim 0.5$;
but as we discuss below, the nature of the bright end of
the supernova luminosity function remains uncertain
and other rare but bright supernova types \cite{TSS,SN2008es,GalYam,quimby}
might also be important at these large redshifts. 

It is important to appreciate that
the most crucial input from future synoptic surveys 
will be the {\it normalization} of the CSNR.
The shape of the CSNR follows from that of 
the star-formation rate due to the very short lifetimes
of all massive star progenitors, and the cosmic star-formation
redshift history is already
relatively well-known out to $z \sim 1$.
However, the CSNR normalization is only known to within $\sim 40\%$.
This will be greatly improved by future synoptic surveys, which
should measure the CSNR to extremely high precision 
at $z \sim 0.3$,
and therefore dramatically reduce the uncertainties
in the CSNR (and hence the DSNB) normalization.

Because our focus is on the interplay between synoptic surveys and
neutrino observations,
we wish to carefully distinguish different outcomes for massive stars and
their resulting optical and neutrino emission.
All collapse events produce neutrinos; 
however, simulations have shown that 
both the amount and energies of the supernova neutrinos 
varies with the mass range of the progenitor stars and 
how they end their lives 
\cite{fryer99,daigne05,sumiyoshi07,horiuchi,sumiyoshi08,nakazato,lunardini09}.
Unfortunately,
there exists great uncertainty
about the fate of massive stars, and
the as-yet unresolved physics of the baryonic explosion mechanism
may well play an important role in determining the
outcomes \cite{buras05,janka06,mezzacappa,thompson,sumiyoshi07}.
Recent work 
suggests that stars below some characteristic mass 
(estimated at $\sim 25 \msol$)
do explode, producing optical supernovae and leaving behind
neutron stars;
on the other hand, 
stars above some mass scale (estimated at $\sim 40 \msol$) 
are expected to collapse directly 
into massive black holes without
optical signals \cite{fryer99,heger,nakazato}.
It is possible that between these regimes, 
a mass range exists (e.g., $25-40 \msol$) 
that would be a gray area where core-collapses
form black holes from fallback while
still being able to display some (perhaps dim) optical signals.

In the following sections, we will refer to those massive stars
that first undergo regular core collapse and bounce
as ``core-collapse'' events, 
whether they ultimately leave behind neutron stars or black holes
formed from fallback.
Those massive stars that collapse directly to
black holes we will refer to as
``direct-collapse'' events.
Events that also produce substantial electromagnetic outbursts
we refer to as ``visible''; those that do not are ``invisible.''
For simplicity, but also following current thinking,
we take visible events to be core-collapse events 
that produce neutron stars and conventional 
(i.e., 1987A-like) neutrino signals.
We take invisible events to be direct-collapse events,
which have a higher-energy neutrino signal \cite{nakazato}.
``Failed'' supernovae should be invisible from our viewpoint,
though some may have weak electromagnetic signals 
that we henceforth ignore
\cite{macfadyen}.

The focus of this paper is to quantify how
the CSNR determination by future synoptic sky surveys will 
improve the DSNB prediction, and to point out some of the science payoff
of this improvement.
After summarizing the DSNB calculations (\S \ref{sect:formalism}),
we present our forecasts for the CSNR measurements by
synoptic sky surveys (\S \ref{sect:csnr}). 
Using these, we show the impact on the DSNB
(\S \ref{sect:dsnb}).
In particular, 
we discuss 
present constraints on 
invisible events, and strategies for
DSNB data to probe the fraction of massive star deaths that
are invisible
(\S \ref{sect:invisible}).
We then switch to a extremely conservative 
viewpoint and discuss the robust lower limit on the DSNB
(\S \ref{sect:lowerbound}).
Conclusions are summarized in \S \ref{sect:conclude}.

\section{DSNB Formalism and Physics Inputs}

\label{sect:formalism}

The neutrino signal from the ensemble of cosmic collapse events
is conceptually simple,
and is given by the line-of-sight integral of
sources out to the cosmic horizon
(more precisely, to the redshift where star formation begins; 
in practice, the result does not change once redshifts of a few are reached).
The well-known result is
\begin{align}
\label{eq:nuflux}
\phi_{\nu}(\enu)
& = 4\pi I_{\nu}(\enu)
 = c \frac{dn_{\nu}}{d\enu} \notag \\
& = c \int_0^\infty 
  (1+z) \ \left|\frac{dt}{dz}\right| \ 
  \csnr(z) \ N_{\nu}[(1+z)\enu]  \ dz, 
\end{align}
where $I_{\nu}(\enu)$ is the neutrino intensity (flux per solid angle)
of cosmic supernova neutrinos with observed energy $\enu$.
Because Earth is transparent to neutrinos, 
detectors see a total (angle-integrated) flux
$\phi_\nu(\enu)$ from the full sky.
Note that neutrinos and their energies are measured individually,
so the intensity and fluxes measures particle number, 
not the energy carried by the particles.
Two source terms, $\csnr$ and $N_{\nu}[(1+z)\enu]$, 
appear in Eq.~\ref{eq:nuflux}.  
$\csnr$ is the cosmic rate of collapse events, 
i.e., the number of collapse events per comoving volume 
per unit time in the rest frame.
Each source, i.e., each collapse event, 
has a neutrino energy spectrum $N_{\nu}(\enu_{\rm emit})$ 
in its emission frame with rest-frame energy
$\enu_{\rm emit} = (1+z) \enu$; the
factor $(1+z)$ accounts for the redshifting of energy
into the observer's frame.
Because we allow for different neutrino energy spectra
for core-collapse (CC) and direct-collapse (DC) events, $N_{\nu}[(1+z)\enu]$ 
can be expressed as
\beq
N_{\nu}[(1+z)\enu]=f_{\rm CC} \, N^{\rm CC}_{\nu}[(1+z)\enu]
			+ f_{\rm DC} \, N^{\rm DC}_{\nu}[(1+z)\enu],
\eeq
where $f_{\rm DC} = \csnr^{\rm DC}/\csnr$ 
and $f_{\rm CC} = 1 - f_{\rm DC}$ are the fractions for
direct-collapse and core-collapse events, respectively;
we assume these  
to be constants independent of time and thus redshift.
Because these fractions are very uncertain, 
below we will consider a range of possible values.
Finally, for the standard $\Lambda$CDM cosmology the
time interval per unit redshift is
\begin{align}
\label{eq:dtdz}
\left|\frac{dt}{dz}\right|
& = \frac{1}{(1+z) H(z)} \notag \\
& = \frac{1}{(1+z) H_0 \sqrt{\Omega_{\rm m}(1+z)^3+\Omega_{\Lambda}}}. 
\end{align}

Equation~\ref{eq:nuflux} shows that
three inputs control the DSNB:
(i) cosmology, via the cosmic line integral and 
parameters;
(ii) supernova neutrino physics, via the source spectrum. 
(iii) astrophysics, via the CSNR. 
Of these, the cosmological inputs
entering via Eq.~\ref{eq:dtdz}
are very well understood 
and their error budget is negligible.
We adopt the standard $\Lambda$CDM model, 
with parameters from the 5-year WMAP data: 
$\Omega_{\rm m}=0.274$, 
$\Omega_{\Lambda}=0.726$,
and $H_0 = 70.5 \ \rm km \ s^{-1} \ Mpc^{-1}$ \cite{wmap5}.
Within this fixed cosmology, DSNB predictions
require knowledge of the source spectra and CSNR.
The purpose of this paper is to forecast the effects of 
future improvements on the source rate,
but to illustrate these we must adopt source
spectra.

Core-collapse neutrino spectra are in principle calculable
from detailed 
supernova simulations, e.g., 
\cite{buras05,janka06,mezzacappa,thompson,sumiyoshi07}.
In practice, it remains quite difficult to simulate supernova
neutrino emission accurately
within realistic explosion models (if they explode at all!)
and certainly it remains computationally prohibitive
to perform such
{\em ab initio} simulations over
wide ranges of supernova progenitors.
Consequently, in DSNB predictions 
different groups have taken different approaches in
estimating neutrino energy source spectra.
Here, we adopt the treatment in the recent DSNB
forecasts of Ref.~\cite{horiuchi}. These authors
approximated the neutrino energy spectra
as Fermi-Dirac distributions with
zero chemical potential:
\beq
\label{eq:nuspectrum}
N_\nu(\enu)
= {\cal E}_{\nu} \frac{120}{7 \pi^4}
  \frac{\enu^2}{T^4_{\nu}}
  (e^{\enu/T_{\nu}}+1)^{-1},
\eeq
where ${\cal E}_{\nu}$ is the total energy 
carried in the electron antineutrino flavor 
and $T_{\nu}$ is the 
effective electron antineutrino temperature. 
Neutrino flavor change effects
are absorbed into the choices of
${\cal E}_{\nu}$ and $T_{\nu}$.
Following Ref.~\cite{horiuchi}, 
we assume the total energy is equally partitioned 
between each neutrino flavor for both 
core-collapse and direct-collapse events,
i.e. ${\cal E}_{\nu} = \etot/6$ for individual neutrino flavor,
where $\etot$ is the total (all-species) energy output.
The variation in neutrino emission from different 
core-collapse progenitor stars
is in general expected to be small because 
neutrinos come from newly-formed neutron stars. 
We adopt
$\etot = 3 \times 10^{53}$ erg per 
core-collapse event. 
Ref.~\cite{horiuchi} finds that
the average temperature after neutrino mixing 
is constrained to lie in the range $T_{\nu} \sim 4 - 8 \ \rm MeV$.
We choose $T_{\nu} = 4 \ \rm MeV$ 
as our benchmark temperature, 
which is close to the empirically-derived spectrum of
SN 1987A \cite{yuksel07}.

For the direct-collapse events,
hydrodynamic simulations show that
the neutrino spectra are sensitive to
the progenitor masses  
and nuclear equation of states, with
models giving
total neutrino energy outputs
ranging from $1.31 \times 10^{53}$ to $5.15 \times 10^{53}$ erg
and different neutrino average energies
ranging from $\epsilon^{\rm avg}_{\nu}$ = 18.6 to 23.6 MeV 
\cite{sumiyoshi07,sumiyoshi08,nakazato}.
We choose the model with higher energy
so it will create a greater difference for comparison.
That is, we take $\etot = 5.2 \times 10^{53}$ erg, and
$T_{\nuebar} = \epsilon^{\rm avg}_{\nu}/3.15 = 7.5$ MeV.

In what follows,
we first take all supernovae to
be core-collapse events (thus visible) as the fiducial case, and then
we will examine the 
impact of the direct collapse (invisible) supernova scenario.
Since the emission from the direct-collapse events 
is taken to be larger, this will increase the DSNB detection rates.
Cosmic supernova neutrinos will be detected 
mainly via inverse beta decay
$\bar{\nu_e} + p \rightarrow n + e^{+}$
interactions with protons in a liquid water or scintillator
detector.
This reaction is endoergic with the threshold energy 
of 1.8 MeV.  
To a good approximation, the nucleon remains at rest, 
so that $\epsilon_{e^+} \simeq \epsilon -\Delta$,
where $\epsilon_{e^+}$ is the positron total energy, 
$\epsilon$ is the $\bar{\nu}_e$ energy, 
and $\Delta = m_n - m_p = 1.295$ MeV.
The expected differential event rate, per unit time and energy, is
\beq
\frac{dR_{\rm detect}}{d\enu}
 = N_p \ \sigma_{\nu p}(\enu) \ \phi_{\nu}(\enu) \ \ .
\label{eq:nudetect}
\eeq
The well-known inverse beta decay
cross section $\sigma_{\nu p}(\enu)$ \cite{vogel,strumia},
taken here at lowest order, 
and which increases with energy roughly as $\enu^2$. 
Thus the event rates give larger weight 
to the high-energy neutrino flux, 
which, as we will see is the regime best probed by supernova surveys.
The total event rate in a detector sensitive to
neutrino energies $\enu$
is thus 
$R = \int^{\epsilon_{\rm max}}_{\epsilon_{\rm min}} dR/d\epsilon \; d\epsilon$.
The factor $N_p$ in Eq.~\ref{eq:nudetect}
gives the number of free protons (those in hydrogen atoms) in the detector;
in our calculations, we use the value
corresponding to 22.5 kton of pure water for Super-K.

\begin{figure}[!htb]
\begin{center}
\includegraphics[width=0.55\textwidth]{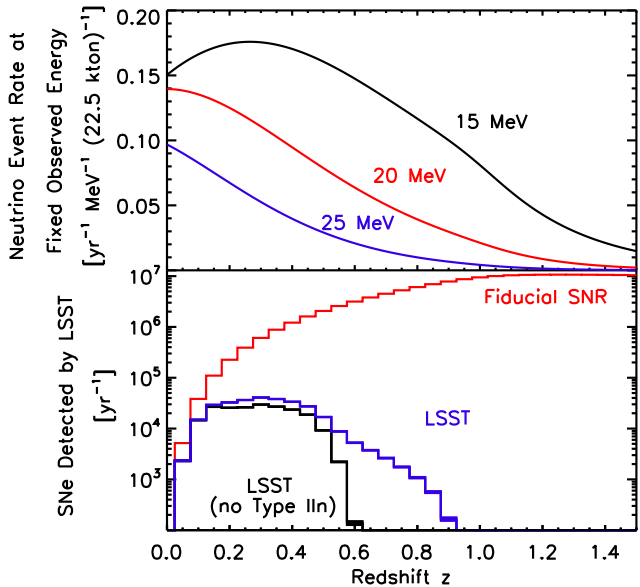}
\end{center}
\caption{
DSNB and synoptic survey redshift distributions. 
{\em Upper Panel}:
The integrand of Eq.~\ref{eq:nuflux} as a function of redshift
for different choices of observed neutrino energies; 
this shows the redshift distribution of sources that
contribute to the DSNB signal at these energies.
Here we assume all the supernovae are core-collapse events,
as defined in \S \ref{sect:intro}.
{\em Bottom Panel}:
The blue curve is the 
supernova detection rate by LSST in $r$-band as a function of redshift,
with survey depth $\mlim = \mag{23}$
and sky coverage of 6.1 sr (20,000 $\rm deg^2$).
The black curve is a more conservative estimation
of the LSST supernova detection rate 
by excluding Type IIn supernovae, which seem to 
be the most luminous based on the small sample of current data.
The red curve is the fiducial supernova rate for comparison, 
which is the full-sky supernova rate without considering 
dust extinction or survey depth.
The curves have bin size $\Delta z$ = 0.05, and 
the band thickness (which are in most cases thinner than the curve width)
represent the statistical uncertainty $1/\sqrt{N}$.
}
\label{fig:nusnr}
\end{figure}

The upper panel of Fig.~\ref{fig:nusnr}
shows the neutrino event rate
-- the integrand of Eq. (1) with $T_{\nu}$ = 4 MeV -- 
with respect to redshift 
at certain fixed observed energies.
Because of redshift, 
neutrinos with low observed energies are more likely to 
come from high redshift supernovae, while neutrinos with high observed 
energies are more likely to come from low redshift supernovae. 

A measurement of DSNB neutrinos and their
energy spectrum will thus provide unique new insights
into the physics of massive-star death.  But for the DSNB
to usefully probe the neutrino emission from supernova interiors,
the cosmic source rates must be known.
It is to this that we now turn.

\section{DSNB Astrophysics Input}

\label{sect:csnr}

The CSNR not only controls the
DSNB flux, but also is of great intrinsic interest,
and has a direct impact on numerous
problems in cosmology and particle astrophysics.
The stellar progenitors of both core-collapse and direct-collapse
events are very short-lived; consequently
the CSNR is closely related to the cosmic {\em star-formation}
rate, which has been intensively studied for the past decade
\cite{csfr}.
From the present epoch back to $z \sim 1$, the cosmic star-formation rate 
increases by an order
of magnitude.  
At higher redshifts, $z \ga 1$, 
the cosmic star-formation rate becomes less certain, 
but the $z \la 1$ regime is responsible
for a large fraction of the observable DSNB signal.
On the other hand, while the {\em shape} of the 
cosmic star-formation rate is relatively secure,
the absolute {\em normalization} remains harder to
pin down. 
Recent estimates using multiwavelength proxies for the 
star-formation rate 
indicate a $\pm 20\%$ uncertainty at $z=0$ and 
a larger uncertainty at higher redshift, 
producing an average of $\pm 40\%$
uncertainty on the DSNB detection rate \cite{horiuchi}.
For the direct supernova rate data reported in Ref.~\cite{horiuchi},
here we adopt a $\pm 40\%$ uncertainty at $z = 0$, double
that on the star-formation rate itself (this should not be
confused with the 40\% above).

Fortunately, a new generation of powerful sky surveys 
are poised to offer a new, high-statistics measure
of the CSNR.
These surveys have wide fields of view and
large collecting areas, in order to produce deep
scans of large portions of the sky.
These synoptic surveys are designed to 
repeatedly scan a large portion of the sky every few nights 
with limiting single-exposure 
magnitudes of $\sim \mag{21}$ to $\sim \mag{24}$, and possibly
deeper in several passbands.
Relatively more modest prototype synoptic surveys
have already been completed, e.g., 
SDSS-II \cite{sdss-sn} and  
SNLS \cite{snls},
or are underway, e.g.,
the Pan-STARRS 1 prototype telescope has
already seen first light \cite{pan-starrs-sne},
and the Palomar Transient Factory already reported 
their first results \cite{ptf}.
Large-scale planned surveys include DES \cite{des}, 
LSST \cite{lsst,tyson}, 
SkyMapper \cite{keller},
and the full-scale Pan-STARRS.

These synoptic surveys will repeatedly scan the sky with revisit times
(``cadences'') of $\sim$ few days.
The cadence timescale is ideally suited for following supernova
light curves and detecting events near maximum brightness.
Indeed,
the SNLS have reported 289 confirmed Type Ia events and
117 confirmed core-collapse supernovae out to $z \sim 0.4$ \cite{snls}.
SDSS-II also reported 403 spectroscopically confirmed events \cite{sdss-sn}
(most of which were Type Ia),
and 15 confirmed Type IIp events 
that are potentially capable of being used as standardized candles
\cite{DAndrea09}.
The Palomar Transient Factory has already found \cite{quimby}
three events which are among the most luminous
core-collapse events ever found, and which appear to be
pulsational pair-instability explosions of ultramassive stars.
Finally, Pan-STARRS 1 has reported its first confirmed supernova
\cite{pan-starrs-sne}.

Note that these surveys are {\em unbiased} in that they cover
a large portion of the sky regions systematically
and thus do not pre-select galaxy types or redshifts or
luminosities for supernova monitoring,
whereas most of the past supernova surveys 
monitored pre-selected galaxies 
so that the results were biased, 
though attempts have been made to correct for that.

Currently, most of the design efforts for
synoptic surveys focus on Type Ia supernovae,
because these events are a crucial 
cosmological distance indicator at large redshifts.
However, the survey requirements for Type Ia supernova
detection
are also well-matched to collapse events,
and therefore surveys that 
are tuned for Type Ia supernovae will 
automatically observe collapse events also.
With their proposed properties, these surveys are expected to
discover $\sim 10^5$ collapse events per year out to redshift $z \sim 1$ 
\cite{young08,lf}.
Due to the large sample size, spectroscopic followup is unfeasible
for most events, so photometric redshifts of
the host galaxies (for which deep co-added fluxes will be available)
or of the supernovae themselves will be needed, just as in the
case of Type Ia events \cite{zheng}.

Lien \& Fields~\cite{lf} give detailed predictions 
for the supernova harvest by synoptic
surveys; here we summarize the key factors important for the DSNB.
Within the 5-color SDSS $ugriz$ bandpass system,
the $r$ and $g$ bands provide the largest supernova harvest, due largely to
high detector efficiency for these wavelengths.
Moreover, distant intrinsically blue collapse events are redshifted
into these bands.
Detection of a supernova is done by differencing 
exposures of the same field of view.
To determine if a transient is a supernova and to
establish its type, one must follow the supernova through
the rise and fall of its light curve.
Consequently the peak flux
must be brighter than the minimum flux
for point source detections,
and following Ref.~\cite{lf} we set a supernova limiting magnitude
$\mlim = m_{\rm lim} - \mag{1}$ that is brighter by $\mag{1}$ than the
single-visit point-source limit $m_{\rm lim}$.
Finally, for a given scan cadence timescale, 
a survey must trade off 
scan area $\oscan$
and exposure depth $\mlim$.
Surveys with large scan area,
such as Pan-STARRS and LSST,
are planned to have survey depth
$\mlim = \mag{23}$.

The blue curve in the lower panel of Fig.~\ref{fig:nusnr} plots 
the expected collapse event rate detected by LSST in $r$-band.
One can see from the plot that in one year, LSST will
have more than 100 supernova detections in all 
$\Delta z = 0.05$ redshift bins out to redshift $z \sim 0.9$,
and for $z \simeq 0.1 - 0.5$, LSST will be able to detect more than
$10^4$ supernovae in each bin.
Ref.~\cite{lf} shows that  
Type IIn supernovae contribute to most of the 
detections for $z \gtrsim 0.5$
based on the luminosity functions provided in Ref.~\cite{richardson}. 
Since this higher end of the detection redshift range
is highly affected by the small sample of Type IIn in Ref.~\cite{richardson},
we also plot the black curve for reference to show
a more conservative estimation
that excludes Type IIn supernovae.
One can see that the detection would reach $z \sim 0.6$
in this case. 
The thickness of the blue and black
curve represent the statistical uncertainty ($1/\sqrt{N}$),
which in most cases are thinner than the curve width because the uncertainty 
is very small due to the large number of supernovae. 
The full-sky fiducial supernova rate based on Ref.~\cite{horiuchi}
is also plotted for comparison. The difference between 
the fiducial supernova rate and the LSST detection rate 
is mainly due to survey depth (magnitude/flux limit), sky coverage
and to a lesser extent dust obscuration. 

A high precision measurement of the CSNR
can therefore be done via direct counting of
the enormous number of collapse events versus redshift.
While a measurement of the CSNR {\em shape} 
will tests the consistency with 
results inferred from other methods, 
such as the star-formation history,
the real power of synoptic surveys will be the 
high-statistics determination of the
CSNR {\em normalization}.  Note that this
can in principle be determined by precision measurement of
the CSNR at a {\em single} redshift bin, where the
counts are largest.
For a large survey like LSST,
this should occur around $z \sim 0.3$, which is set by the
tradeoff of survey volume and limiting magnitudes
\cite{lf}.
In general,
LSST is expected to probe 
the CSNR
out to redshift $z \sim 0.9$ to 
$1/\sqrt{N} \sim 10\%$ statistical precision within one year
of observation.

As mentioned earlier,
detections in the $z \sim 0.5-0.9$ range 
will be dominated by the most luminous core-collapse events.
In a study of the core-collapse luminosity function
based on relatively sparse and inhomogeneously taken data,
the relatively rare Type IIn events were found to be the
most intrinsically luminous \cite{richardson};
and ultraluminous Type IIn events have been found
\cite{SNIIn}.
Recent observations, including those by the synoptic Palomar Transient
Factory and by ROTSE-III/Texas Supernova Search,
show that other core-collapse types can also
lead to ultraluminous explosions;  of these,
the newly-discovered pair-instability outbursts
are particularly intriguing and encouraging 
because this entire class of events
has likely gone unnoticed until now
\cite{TSS,SN2008es,GalYam, quimby}. 
There is clearly much more to be learned about 
about the bright end of the supernova
luminosity function. 
As more data of these ultraluminous events become available, 
the redshift reach of synoptic surveys 
will come into a much better focus.

\section{Impact of Synoptic Surveys on the DSNB}

\label{sect:dsnb}

We are now in a position to assess the synoptic survey impact on the
DSNB. 
Our viewpoint
is to envision the situation several years from now,
when synoptic surveys have been running in earnest, and 
when the DSNB signal has been at last detected.
Of course, real surveys will miss 
core-collapse events for a variety of reasons, yet following 
Ref.~\cite{lf} we believe there is good reason
to expect that these losses can be calibrated,
empirically or semi-empirically, and thus
the absolute CSNR can be obtained 
out to $z\la 1$;
this should verify the
already well-determined shape of the cosmic star-formation rate
in this regime.
Furthermore, surveys will definitely measure 
the low-redshift {\em normalization}
of the CSNR to high precision via {\em direct counting}.  

To be sure, it will be far from trivial to arrive at the understanding
we presuppose.  There will be formidable astrophysical challenges
in extracting from survey data the supernova properties of interest,
most importantly the event type, redshifts, and obscuration;
less crucially for our purposes one would like as well the
intrinsic luminosity.
Ref.~\cite{lf} discusses some reasons for 
optimism in the face of these challenges,
and we also remind the reader that these issues are crucial
not only for studies of the DSNB but also are central for other
key topics in astrophysics and cosmology.
Most notably, the problems of obtaining supernova type, redshift,
and obscuration are at least as pressing (and in some
respects more challenging) when one uses
supernovae as cosmological distance indicators
and thus as probes of dark energy.
Put differently, if survey supernovae are understood well enough to do
dark energy cosmology, then we expect that the 
star-formation rate should  be well-understood
enough to give the DSNB source history out to $z \sim 1$,
and the CSNR normalization to high precision.

We now explore the impact of a CSNR determination of this kind.
That is, we assume that one can use synoptic surveys to
infer the absolute normalization and shape of the
CSNR out to some redshift $z_{\rm max}$.
In particular, Ref.~\cite{lf} showed that
all core-collapse types should be visible out to $z_{\rm max} \ga 0.5$,
and the very bright Type IIn events should extend to
$z_{\rm max} \ga 1$ \cite{SNIIn,cooke}.
Thus we will take the CSNR {\em shape} 
to be directly known from surveys to $z = 1$,
and following Ref.~\cite{lf} we assume that the
{\em normalization} will be very well-determined 
statistically, and so we will anticipate a measurement
good to $\delta \csnr/\csnr = 5\%$; this error would be
dominated by systematic uncertainties
at the most relevant redshifts.

Referring again to Fig.~\ref{fig:nusnr}, we compare the redshift reach 
of synoptic surveys with the redshift distribution of the DSNB sources.  
We see that the two are well matched. 
That is, within the 
detection energy range ($\sim 10 - 26$ MeV positron energy), 
the neutrino sources peak within the redshift range of 
upcoming supernova surveys.
Quantitatively, 
the detection rate
is about 1.8 neutrinos/year within
the detector energy range of 10 -- 26 MeV positron energy
for neutrinos from all redshifts 
(i.e., $z_{\rm max}=6$).
Of this total rate,
events within redshift $z=1$
contribute 1.5 (87\%) neutrinos/year,
and events within redshift $z=0.5$
contribute 1.0 (54\%) neutrinos/year.
Our results are in good agreement with 
the numbers shown in Ref.~\cite{horiuchi} and \cite{ando}.
Therefore 
a large fraction of the observable neutrinos 
come from events
within $z \sim 1$, 
which is about the same redshift
range as the upcoming supernova surveys.

\begin{figure}[!htb]
\begin{center}
\includegraphics[width=0.55\textwidth]{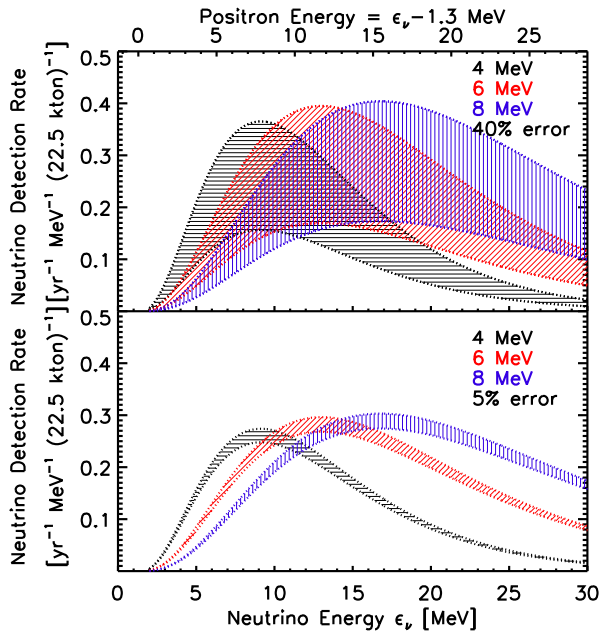}
\end{center}
\caption{
{\em Upper Panel}:
Neutrino detection rate as a function of neutrino observed energy, 
with different neutrino effective temperatures
are plotted for comparison. The band thickness of the curves 
represent a $\delta \csnr/\csnr$ = 40\% uncertainty 
in the current CSNR normalization.
{\em Bottom Panel}:
Same as the upper panel, 
but with a 5\% normalization uncertainty instead, which 
is the uncertainty expected from upcoming supernova surveys with 
one year observations.
}
\label{fig:nudetecterror}
\end{figure}

We thus see that using supernovae to directly infer the CSNR
allows us to robustly predict a large fraction
of the detectable neutrino events. A high precision measurement of the CSNR 
would therefore put a better constraint on the 
DSNB flux, which encodes knowledge of supernova neutrino physics.
For example, one would then be able to 
distinguish the difference between neutrino models
with different effective temperatures, as demonstrated 
in Fig.~\ref{fig:nudetecterror}.

Figure~\ref{fig:nudetecterror} plots the neutrino detection rates 
estimated based on models with different neutrino effective temperatures
($T_{\nu}$ = 4, 6, 8 MeV, respectively) versus neutrino energy 
in the observer's frame. 
The upper panel shows the current $\delta \csnr/\csnr$ = 40\% 
uncertainty 
in the cosmic supernova rate normalization. The bottom panel shows 
the future normalization uncertainty of $\delta \csnr/\csnr$ = 5\%
(dominated by systematics), 
which would be achieved
within one year observation of the upcoming supernova surveys.
One can see that
it is not easy to distinguish different neutrino models with 
the current 40\% uncertainty. However, with
a future 5\% precision, it would be certainly possible 
to distinguish the differences between each models 
and therefore provide a way to study supernova neutrino physics
by combining neutrino detections and supernova surveys. 
 
Moreover,
after several years of exposure, one might hope to
attain statistics sufficient to measure the
{\em difference} between the observed flux
and the contributions from lower-redshift
epochs sampled by survey supernovae.
This difference
encodes a wealth of interesting physics and astrophysics.

\section{Invisible Supernovae Revealed} 

\label{sect:invisible}

The most dramatic possibility for a mismatch between the neutrino and optical
supernova measures would reflect a real lack of optical explosions
due to ``invisible'' supernovae. 
As mentioned in Section~\ref{sect:formalism},
even in the context of conventional models
there is a great uncertainty about whether 
stars with masses between 25 to 40 $\msol$ explode or not.
A Salpeter 
initial mass function $dN_\star/dm \propto m^{-2.35}$ \cite{salpeter},
dictates that for collapse events in the $8-100 \msol$ range,
$\sim$ 90\% are core-collapse events (masses $\la 40 \msol$), 
which in our assumption make optically luminous 
explosions even for those that form black holes from fallback,
and $\sim$ 10\% are direct-collapse events ($\ga 40 \msol$) 
that are optically invisible, but have larger neutrino emission
with greater total energy $\etot$ and 
higher neutrino average temperature $T_{\nu}$.
A relatively conservative case,
which has recently been studied by Lunardini \cite{lunardini09},
would then assume that around 10\% of collapse events failed to explode,
hence one would expect that 
the neutrino flux from neutrino detectors 
would at least be $\sim 10\%$ higher than
neutrino flux from supernova surveys.   

However, there remain large uncertainties in our {\em qualitative}
understanding of massive star death, not to mention even larger
quantitative uncertainties in neutrino and photon outputs.
If, as expected, the neutrino emission
is larger for these events than for ordinary supernovae,
then the signal increase in the detectors
can be significantly larger
\cite{daigne05,lunardini06,sumiyoshi07,horiuchi,sumiyoshi08,nakazato,lunardini09}.
Given these substantial uncertainties
it is entirely possible that the invisible fraction is much higher than 10\%.
For example, one possible scenario is that 
supernovae that form black holes from fallback 
might actually belong to the invisible events category.
Ref.~\cite{fryer09} predicts the light curves
of these fallback events with
peak magnitudes 
around $V = -13$ to $-15$,
which correspond to luminosities several orders of magnitude
lower than ordinary core-collapse events.
These authors also suggest that 
the total neutrino emission from the 
fallback events can be larger than normal supernovae \cite{fryer07}.
Thus if we treat the fallback supernovae as 
invisible events with larger neutrino emission,
the invisible fraction will be higher than current
estimates would suggests \cite{lunardini09}.
Therefore we will take the invisible fraction as 
an {\it a priori} free parameter, 
and explore constraints based on neutrinos and other observables.

\begin{figure}[!htb]
\begin{center}
\includegraphics[width=0.55\textwidth]{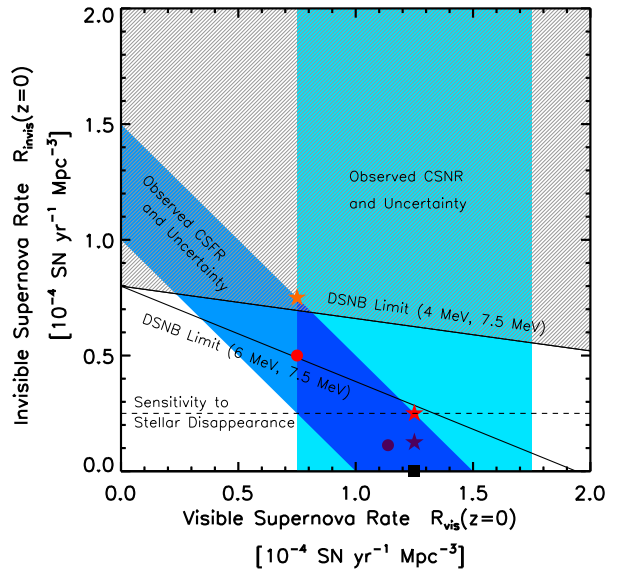}
\end{center}
\caption{
Summary of current and future constraints on the invisible supernova 
rate $\invis$ 
(i.e. the direct-collapse event rate in our assumption) and 
the visible supernova rate $\vis$
(i.e. the core-collapse event rate in our assumption).
Blue regions are those allowed by current 
observed cosmic star-formation rate (CSFR) and CSNR
and their uncertainties.
The grey-lined region is disallowed based on 
the non-detection of the DSNB by Super-K 
with the assumption that $T_{\nu}$ = 4 MeV for visible events 
and 7.5 MeV (and also a higher total energy) for invisible events.
Another DSNB limit with $T_{\nu}$ = 6 MeV instead of 4 MeV 
is also plotted for comparison.
The horizontal dashed line shows the sensitivity to stellar disappearance,
which will directly probe the invisible supernova rate \cite{kochanek}.
Note circles explored in Fig.~\ref{fig:nucomerror} 
and stars in Fig.~\ref{fig:nucomadderror}.
Square marks a baseline shown in Fig.~\ref{fig:nucomerror} and \ref{fig:nucomadderror}.
}
\label{fig:nulimit}
\end{figure}

Fig.~\ref{fig:nulimit} shows several constraints
on the visible supernova rate $\vis$ 
and invisible supernova rate $\invis$ 
at $z=0$.
These constraints are estimated based on 
current data with the assumption that the {\em shape }
of the CSNR is known, and
we adopt the fiducial model described in Ref.~\cite{horiuchi}.
Blue regions in the plot represent 
the allowed regions; the gray region represents 
the explicit exclusion from the non-observation of neutrinos;
and white regions represent areas that are disallowed implicitly, that is, 
they lie outside of current allow regions 
but are not banned directly based on 
current limits.  

One way to constrain $\csnr$ is using
the current {\it observed cosmic star-formation rate}.
The ratio of massive star counts per unit mass into all stars
depends only on the choice of initial mass function;
we take this ratio to be $0.007/\msol$ assuming 
the Salpeter Initial Mass Function (IMF)
\cite{salpeter}.
With the uncertainty 
$\sim 20\%$ in the cosmic star-formation rate normalization \cite{horiuchi},
the upper and lower limit of current star formation rate at $z=0$ correspond
to $\csnr(0) =1.25 \pm 0.25 \times 10^{-4} \, \rm yr^{-1} \, Mpc^{-3}$.
respectively, which
set the darker blue region in Fig.~\ref{fig:nulimit}.
Also, the present {\it observed CSNR} 
with $\sim 40\%$ uncertainty in its normalization is 
plotted as the light-blue region in Fig.~\ref{fig:nulimit},
which correspond to the value of 
$\csnr(0) = 1.25 \pm 0.50  \times 10^{-4} \, \rm yr^{-1} \, Mpc^{-3}$.

The {\it DSNB limit} in Fig.~\ref{fig:nulimit} 
shows the constraint estimated 
from the current non-detection of the 
supernova neutrino background, which sets  
an upper bound of the total core-collapse supernova rate 
$\csnr = \vis+\invis$.
Ref.~\cite{yuksel05}
points out that the upper limit on the neutrino flux 
set by Super-K in 2003 corresponds to an upper limit of 2 events 
per year for a 22.5 kton detector in the energy range of 18 -- 26 MeV 
(see also Ref.~\cite{lunardini_peres}
for the temperature dependence 
of the Super-K limits in terms of flux instead of event rate).
For the benchmark $T_{\nu}$ = 4 MeV case, this limit
allows the current $\csnr$ to be 4.7 times bigger 
than current fiducial value if we assume all 
neutrino emission comes from visible events.
On the other hand, the Super-K limit implies a current
$\csnr$ that is 0.64 times smaller than our fiducial value
if all neutrino emission comes from invisible events. 
Note that these two factors are not 
the same because there is more neutrino emission per invisible event.

The DSNB constraint has substantial uncertainties 
from both the visible and invisible supernova contributions.
The neutrino emission from visible events depends on the neutrino
emission spectrum, i.e., temperature.
To illustrate how this would change the DSNB limit, 
we also plotted the DSNB limit when assuming visible events have 
$T_{\nu}$ = 6 MeV instead of 4 MeV. The 6 MeV line intersects 
the $\vis$ axis
at 1.9 instead of 5.8 for the 4 MeV line. 
While the uncertainty in the neutrino emission from visible events would 
affect where the DSNB limit intersect with the $\vis$ axis, the
uncertainty in the neutrino emission from invisible events would change
where the limit intersects with the $\invis$ axis. 
In this paper we adopt the highest-energy case for 
the neutrino emission from invisible events; however, 
if we choose the lowest-energy case in Ref.~\cite{nakazato}, 
then the limit would intersect with the $\invis$ axis at 4.6 
and the whole region shown in Fig.~\ref{fig:nulimit} would be 
allowed by this limit and thus would give a weaker constraint.

In addition to constraints based on current observational data, 
Kochanek et al.~proposed new method of probing invisible supernovae 
\cite{kochanek}.
These authors suggested monitoring a million supergiants,
in galaxies within 10 Mpc. 
Because the supergiant phase lasts $\sim 10^6$ years,
every year about one monitored supergiant will end its life. 
While some events 
will result in an ordinary optically bright supernovae, 
if any events lack optical outbursts
-- and are thus invisible by our definition -- they will
simply {\it disappear} in sight.  
Considering that the local cosmic star-formation rate 
is about two times higher 
than the cosmic average, 
the lowest invisible event rate 
that predicts one disappearing event in the proposed five years observation 
is around 
$\invis = 0.25 \times 10^{-4} \, \rm yr^{-1} \, Mpc^{-3}$.
This line is shown as the horizontal line 
labeled as {\em sensitivity to stellar disappearance} 
in Fig.~\ref{fig:nulimit}.

Despite the preliminary nature of 
some of the constraints in Fig ~\ref{fig:nulimit}, 
several interesting trends already emerge.    
The allowed region for invisible supernovae is nonzero, 
but it is bounded and cannot be arbitrarily large. 
Future observations will severely restrict the allowed region 
for visible supernovae.
Obviously, the mere demonstration that $\invis$ 
is nonzero would immediately offer novel 
and unique insight into supernova physics.  
Moreover, any quantitative determination of 
the absolute value of $\invis$ 
or the ratio $\invis/\vis$ 
would give detailed insight into the explosion mechanism 
over the full range of core-collapse events.

\begin{figure}[!htb]
\begin{center}
\includegraphics[width=0.55\textwidth]{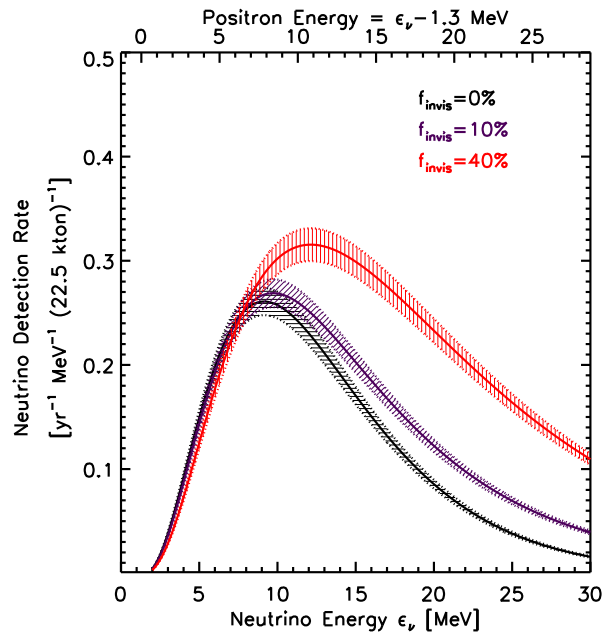}
\end{center}
\caption{
One year neutrino detection as a function of neutrino observed energy.
Three different fractions of the invisible events are plotted 
with $\csnr$ fixed to the fiducial number.
Curves with different colors 
correspond to the square/circles with the same color
in Fig.~\ref{fig:nulimit}.
The band thickness of the curves represent 5\% uncertainty expected from 
upcoming supernova surveys. 
}
\label{fig:nucomerror}
\end{figure}

Also, Fig.~\ref{fig:nulimit} allows a larger 
invisible fraction than the $f_{\rm invis}=10\%$ 
predicted from current theory. 
We marked several possible invisible fractions 
that we will discuss more in the figures below.
The square represents a baseline, 
with invisible fraction $f_{\rm invis} = 0\%$.
Circles mark possible $f_{\rm invis}$ assuming 
the total CSNR is fixed to the fiducial number of
$\csnr=1.25 \times 10^{-4} \, \rm yr^{-1} \, Mpc^{-3}$.
The purple circle is the conservative case with $f_{\rm invis}=10\%$,
and red circle marked the highest invisible fraction
($f_{\rm invis}=40\%$)
one can reach with $\csnr$ fixed.
The corresponding changes in the DSNB detection 
are shown in Figure~\ref{fig:nucomerror}, where we 
see that when error in $\vis$ drops to 5\%, 
it will become possible to tell the difference between these three
cases in the detectable neutrino energy range.
The energy dependence of the fraction traces back to the higher
energy of the neutrino flux from black hole forming supernovae.
Therefore invisible events contribute a larger fraction of
the neutrino flux at higher neutrino energy.

Another set of key points in Fig.~\ref{fig:nulimit}
are marked with stars. In choosing these points, we allow
for the uncertainties in $\vis$ in order
to explore even higher possible $f_{\rm invis}$ values
while staying within current limits.
If the visible event rate is fixed to
the fiducial number of
$\vis = 1.25 \times 10^{-4} \, \rm yr^{-1} \, Mpc^{-3}$,
then 
the purple star marks the point with $f_{\rm invis}=10\%$
adding to current fiducial $\vis$,
and the red star marks the point with $f_{\rm invis}=17\%$, 
which is the highest $f_{\rm invis}$
one can reach with $\vis$ fixed. 
However, the visible event rate is quite uncertain 
and could fall substantially below our fiducial value.  
Including this uncertainty,
the highest $f_{\rm invis}$
that is allowed by current limit is around the point marked 
by the orange star with $f_{\rm invis}=50\%$.  
Note that this point seems to lie just outside the DSNB constraint,
however, one should keep in mind that the DSNB constraint 
is very sensible to theoretical assumption of the supernova 
neutrino emission and hence has its own uncertainty, 
as discussed earlier.

\begin{figure}[!htb]
\begin{center}
\includegraphics[width=0.55\textwidth]{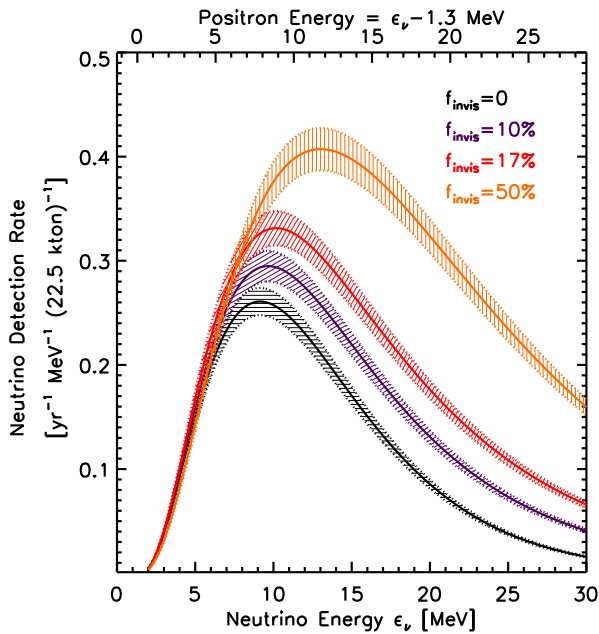}
\end{center}
\caption{
Similar to Fig.~\ref{fig:nucomerror}. However, we allow 
larger numbers for $\csnr$. 
Curves with different colors correspond to square/stars
in Fig.~\ref{fig:nulimit}.
The band thickness of the curves represent 5\% uncertainty expected from
upcoming supernova surveys.
}
\label{fig:nucomadderror}
\end{figure}

The DSNB detections corresponding to the points marked by stars 
are shown in
Figure~\ref{fig:nucomadderror}.
Note that the black curve with $f_{\rm invis}=0\%$ 
represents the neutrino detections from 
the visible events, and thus is the one that would 
be estimated by supernova surveys;
the purple and red curves include different fractions
of invisible events on top of the visible events,
which represent 
those that would be detected by neutrino detectors.
Therefore Fig.~\ref{fig:nucomadderror} illustrates 
how the differences between DSNB
from neutrino detectors and supernova surveys would
encode information of the fraction of invisible events.
Again, the band thickness in this figure indicates the 
expected 5\% uncertainty in $\vis$,
and it is clear that these three cases will be distinguishable.
The DSNB detections for the very extreme case with $f_{\rm invis}=50\%$
is plotted as the orange curve for comparison.

A 50\% invisible event fraction would lead to 
a significant difference between flux from neutrino detectors and 
supernova surveys. 
We find that neutrinos due to invisible events
within $z \sim 1$ 
would contribute around 75\% of the event rate
in the detectable energy range.
For comparison, we expect the neutrinos 
associated with dust-obscured supernovae
to be about $\sim 20\%$ of the signal.
Thus, if the invisible event fraction approaches current limits,
the neutrino census of supernovae should be able to rapidly and strongly 
point to the large contribution from these events. 
Additionally,
an invisible event fraction of 50\% 
could push the mass limit of the 
direct-collapse events to as low as $\sim$ 14 $\msol$
with the Salpeter IMF. However, theories 
about supernova progenitors remain quite uncertain
and therefore the lower mass limit implied by the invisible fraction
is also not necessarily well-defined.
Once the upcoming surveys put better constraints on 
the invisible fraction, 
one can hope to learn more about the 
mass limit of direct-collapse events.

\section{Astrophysical Challenges and Payoffs}

\label{sect:lowerbound}

Our discussion until now has taken a point of view
that by the time synoptic surveys are well under way,
the loss of supernova detections from 
dust and survey depth 
can be corrected, either using the survey data themselves
or from followup observations.
In this section, we change our
viewpoint from this optimistic, wide-ranging anticipation of
future progress to a more restricted focus
on the power of the survey-detected supernovae alone.

For real surveys, 
some of the collapse events must be lost from detection
mainly due to three factors: survey limiting magnitude, 
dust obscuration, and the invisible events without optical explosions.
On the other hand, neutrino detection will be unaffected 
by any of these issues.
Therefore, neutrino flux from neutrino detectors
should exceed that estimated from supernova surveys.

Supernova surveys thus provide a totally empirical, model-independent  
method
to estimate the
{\em extreme lower limit} to the DSNB
by simply adding up        
the neutrino contribution from each supernova detected.
The resulting lower bound
to the DSNB flux is
\beq
\label{eq:srnfluxsurvey}
\phi_{\nu}^{\rm min}(\enu)
\equiv \phi_{\nu}^{\rm survey}(\enu)=\frac{4\pi}{\oscan \Delta t} \ 
\sum^{\rm survey \ SNe}_{i=1} \frac{N_{\nu}[(1+z_i)\enu]}{4\pi D_L(z_i)^2}
\eeq
where each term in the sum is the flux contributed by 
each supernova observed in the survey, and the
prefactor includes a correction for the fraction
$\oscan/4\pi$ of the sky covered by the survey.
The fluxes depend on the luminosity distance $D_L(z)$,
which is fixed by
precisely known cosmological parameters.
Notice that in this equation, only the neutrino energy spectrum 
$N_{\nu}[(1+z_i)\enu$]
depends on supernova and neutrino physics. 

This 
``what you see is what you get'' approach
is robust but conservative.
Namely, the result
$\phi_{\nu}^{\rm survey}(\enu)$ will be 
an extreme lower bound for the DSNB flux.
More detailed and quantitative discussion 
can be found in Appendix~\ref{app:A}.

Once the DSNB is detected, the
{\em difference} between the detected flux
and the survey-based lower bound
provides a unique measure of the events
unseen by surveys.
For example, it is conceivable
that the survey predictions could
{\em exceed} the DSNB detection (or upper limit!).
This result would be very surprising and thus extremely tantalizing,
as it would challenge
our assumptions related to supernova physics 
and neutrino physics.
In other words, this would mean that one or both terms 
in Eq.~\ref{eq:srnfluxsurvey},
the luminosity distance $D_L$ and/or the supernova
neutrino emission spectrum $N_{\nu}[(1+z_i)\enu]$, 
might be wrong.
But the physics behind $D_L$ rests on well-established
Friedmann-Robertson-Walker cosmology, 
and depends only on well-determined cosmological parameters.
Thus a ``DSNB deficit'' would much
more likely point to problems in the supernova
emission spectrum $N_{\nu}(\enu)$.
Therefore, if the lower bound estimation $\phi_{\nu}^{\rm survey}(\epsilon)$ 
turn out to be higher than the actual neutrino detections, 
we would be driven to rethink 
supernova neutrinos in a way to substantially reduce the
observable signal.

The more likely and certainly more conventional expectation
is that when the DSNB is detected, its flux will be
higher than the supernova survey lower bound
$\phi_{\nu}^{\rm survey}(\epsilon)$.
In this case, the {\em sign} of the difference would
be unsurprising, but the {\em magnitude} of the
detected versus survey excess would still 
encode valuable new information,
such as the invisible fraction 
as discussed in the previous section.

One might also hope for the possibility to combine 
survey supernovae and the DSNB to probe events that
{\em are} optically visible but are lost due to dust obscuration;
this could give insight into the nature and evolution of cosmic
dust.
To see how $\phi^{\rm survey}_{\nu}$ would change with different 
dust models,
we examine with two extreme cases:
(1)  model with extremely low dust obscuration by assuming constant dust obscuration as those at local universe
mentioned in Ref.~\cite{mannucci}; and 
(2) a model with very high dust obscuration by doubling the dust evolution with redshift compares to the
model suggested in Ref.~\cite{mannucci}.
We find that
with $\mlim=23$, the neutrino detection rate estimated from
uncorrected 
supernova surveys changes by only $\sim 7\%$ when comparing these two models.
That is, dust models (1) and (2) give
0.34 to 0.31 events per year, respectively. 
Therefore
the neutrino detection rate estimated from
supernova survey is insensitive to the dust models
and hence it will be difficult to use 
the DSNB to distinguish different dust models
with the expected survey precisions.

\section{Conclusions}

\label{sect:conclude}

With the next generation synoptic surveys coming online,
a high precision measurement of the CSNR 
via {\it direct counting} will be achieved,
and thus greatly reduce the uncertainty in the DSNB to a few percent.
An interlocking set of strategies suggest themselves,
by which one can leverage survey supernovae and
the DSNB to probe neutrino physics as well as
the astrophysics of cosmic supernovae.
For example, the high-precision DSNB prediction based on
supernova surveys 
would be able to distinguish supernova neutrino models
with different neutrino temperatures.

As we have shown, the $z \la 1$ DSNB contribution
comprises most of signal at high energy $\ga 10$ MeV,
and so a comparison of the
high-energy predictions and observations
would measure
the amount of events unseen by surveys.
One of the exciting possibilities is 
using the DSNB to probe the fraction of invisible events.
With the current uncertainties,
the observed cosmic star-formation rates and the CSNR 
already suggests possible ranges 
for the invisible fraction.
Indeed, limits from present observables allows 
a substantial invisible events to up to $\sim 50\%$,
which is much higher than the fraction suggests by current 
supernova theories ($\sim 10\%$).
Once the upcoming synoptic surveys begin 
and provide 
high precisions on the CSNR and the cosmic star-formation rate, 
one can hope to reveal the 
fraction of invisible events.

The current non-detection of the DSNB flux 
also limits the total supernova rate.
However, this limit is sensitive 
to the theoretical assumptions of the 
total neutrino energy $\etot$ and 
neutrino temperature $T_{\nu}$.
Therefore the high precision of the DSNB prediction
inferred from upcoming supernova surveys
will make this limit stronger 
by providing knowledge 
of supernova neutrino physics.

While it is unknown whether and to what degree truly invisible supernovae
occur, it is certain that 
survey depth and dust obscuration 
will also hide supernovae from detections.
To interpret the supernova data physically demands that we distinguish
between these factors.
While the loss from survey depth is likely to be corrected by 
knowledge of supernova luminosity function, 
to entangle the degeneracy between dust obscuration and invisible events 
will be challenging.
However, we believe it is not impossible to discriminate the two.
For example,
there are observables across multiple wavelengths that
can be used to estimate dust extinction.
If we can constrain the amount of dust to a higher precision
by combining all different ways of measuring dust,
then the dust effects can be modeled out \footnote{A possible cross-check here are Type Ia
events.  These are due to an older stellar population than core-collapse events
and thus should not be preferentially obscured in their immediate locations;
however, those in spiral galaxies will still suffer obscuration 
by host-galaxy disk material 
that happens to lie along the line of sight.  Thus Type Ia obscuration
and reddening should set lower limits to the effects suffered by core-collapse events.}.
Hence, the only left main unknown would be the 
fraction of invisible events and we could 
learn this fraction by comparing the 
neutrino flux from neutrino detectors and supernova surveys.

On the other hand, 
even without any extrapolations to 
the original observational data,
precision measurement of 
the CSNR will be achieved by 
upcoming surveys,
and thus will
infer a robust lower limit
of the DSNB flux by 
simply adding up the 
neutrino contribution from each supernova.

We conclude by again underscoring the happy accidents that
large-scale synoptic sky surveys will come online
just at the time that large neutrino experiments should
first discover the DSNB, and that the redshift reach of the
two are comparable.
By exploiting the interconnections among the results from
these observatories, we have a real hope of shedding new
light into particle physics and particle astrophysics.

\begin{acknowledgments}
We are pleased to thank Avishay Gal-Yam 
and Jim Rich for enlightening discussion of
supernova discovery by synoptic surveys and
the challenges and opportunities these present.
We are grateful to the anonymous referee for
helpful comments that have improved this paper.
We would also like to thank 
the Theoretical Physics Institute at the University of 
Minnesota, and to the Goddard Space Flight Center
for their hospitality while some of this work was done.
J.F.B. was supported by NSF CAREER Grant PHY-0547102.
\end{acknowledgments}

\appendix

\section{Surveys Set a Model-Independent Lower Bound to the DSNB}

\label{app:A}

\begin{figure}[!h]
\begin{center}
\includegraphics[width=0.45\textwidth]{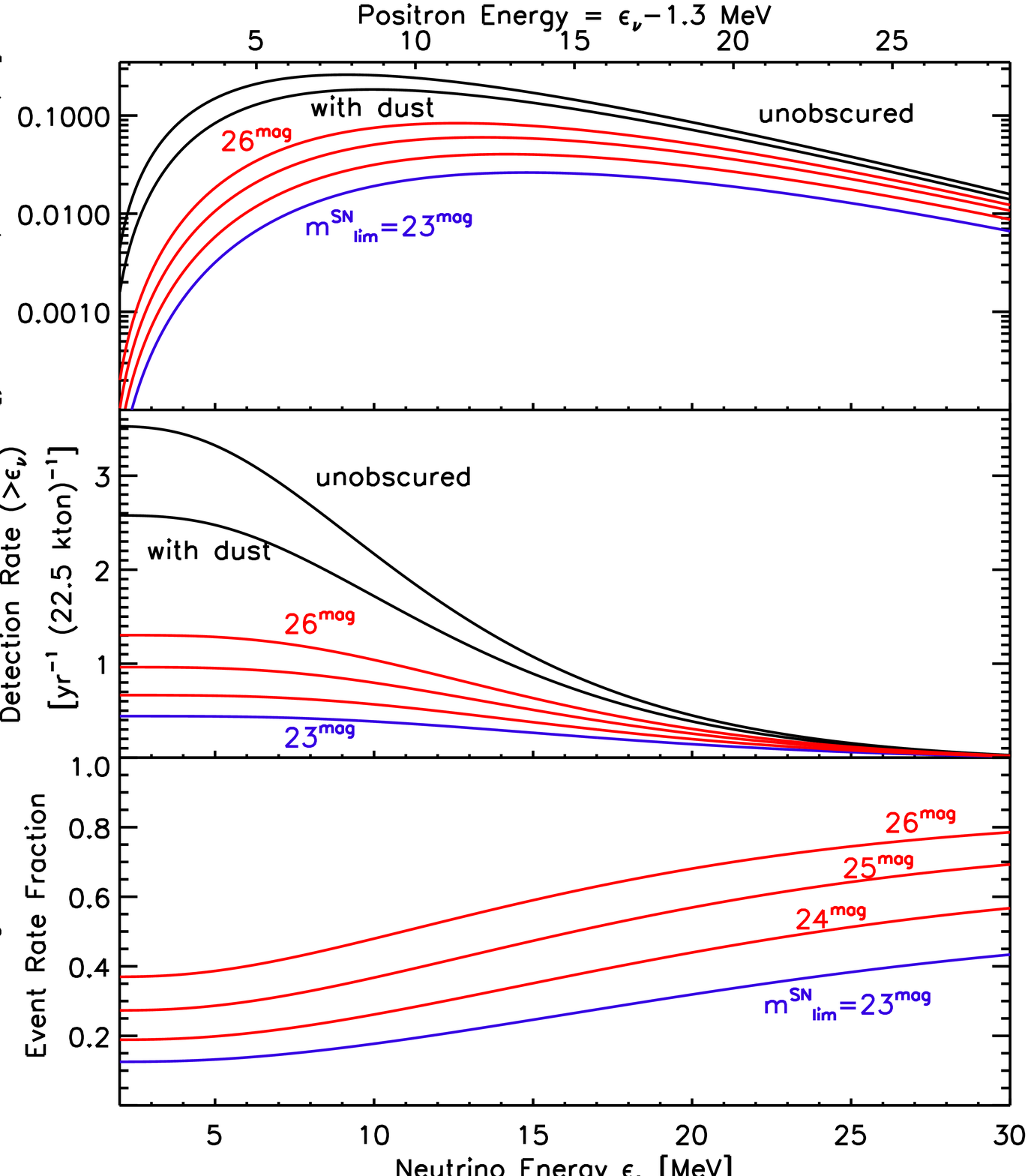}
\end{center}
\caption{
{\em Upper Panel}: Extreme {\em lower bounds} to the
DSNB detection rate
obtained by summing supernovae observed by surveys with different
limiting magnitudes, the blue curve is the limiting magnitude proposed
by LSST and Pan-STARRS. The blue and red curves represent
a lower limit because they apply no correction for
supernovae that are too dim or too obscured to be seen
in surveys.
The two black curves are shown for comparison:
Top black curve is the DSNB flux from
all core-collapse events in the universe out to redshift z $\sim$ 6.
Second top black curve is the DSNB flux from core-collapse events
after considering dust obscuration but with infinite
survey limiting magnitude.
Results assume $T_{\nu}= 4$ MeV.
{\em Middle Panel}: The integrated DSNB detection rate,
i.e. the detection rate above a certain
antineutrino energy and integrated out to
$\epsilon_{\nu}$ = 30 MeV.
The colors indicate the same features as in
the top plot.
{\em Lower Panel}: The fraction of the DSNB detection rate from
the observed core-collapse events
over those from the total collapse events.
That is, a middle-panel red/blue curve
divided by the highest black curve.
Note that in this figure the x-axis starts
at 2 MeV because no events can be
detected below
the threshold energy of 1.8 MeV.}
\label{fig:nuobsr}
\end{figure}

As mentioned in Section~\ref{sect:lowerbound}, a conservative and 
robust lower bound of the DSNB flux can be 
predicted by upcoming supernova surveys. 
Figure~\ref{fig:nuobsr} shows
our estimations for the {\em lower bounds} to
the neutrino flux inferred
from the core-collapse events detected in the $r$-band by
a synoptic survey.
We keep $\oscan$ fixed for simplicity,
but show dependence on $\mlim$ to illustrate
the sensitivity to this parameter.
Planned surveys have sophisticated scan strategies using
a variety of cadences;
for reference, the largest scan areas
of Pan-STARRS and LSST are planned to have
a sensitivity of $\mlim \approx \mag{23}$
in the bandpasses of interest. 

The upper panel shows the predicted neutrino detection rate
from the observed core-collapse events
versus neutrino energy.
Results for the neutrino detection rate
from core-collapse events observed with different limiting magnitude
(from $\mlim = \mag{23} - \mag{26}$)
are plotted. Additionally,
the highest black curve plots
the detection rate from {\em all} core-collapse events
within the horizon (i.e., with no limiting magnitude applied) 
for comparison.
The second highest black curve,
also shows the detection rate for
infinite survey
limiting magnitude, but shows an estimate of the effect
of dust extinction in the host galaxy.
The middle panel shows the integrated neutrino detection rate
$\phi_\nu^{\rm survey}(> \enu)$ above
energy $\enu$.
In other words,
this is the energy-integrated version of the upper panel.
The lowest panel shows the fraction of the neutrino detection rate
from the observed supernovae over the events from all supernovae in the 
universe,
that is, the corresponding middle-panel
red/blue curve divided by the highest black curve.

The difference between the two black curves in Fig.~\ref{fig:nuobsr}
gives an indication of the neutrino contribution from
dust-obscured supernovae. 
We see that an even larger effect is 
the loss of supernovae due to finite survey limiting
magnitude.  
Note that 
when adding dust effects and limiting magnitudes,
the reductions of detection rates
are more severe at low neutrino energies. This is because
observed neutrinos are redshifted, and as a result,
a larger portion of low-energy neutrinos
come from higher redshift
where dust obscuration
is more severe and supernova apparent magnitudes
are dimmer because of larger distances.

The observability of this energy dependence 
is to be understood in the context of
the energy threshold of the neutrino detectors.
For example, Super-K in its present
form can discriminate from atmospheric backgrounds, and 
thus detect, cosmic neutrinos in the $\sim 18-26$
MeV range. If Super-K is enhanced with gadolinium \cite{beacom03},
background rejection would be sufficiently improved in the 10 -- 18 MeV range
to open this crucial window onto the DSNB.

One sees more directly from the lower panel what 
portion of the total neutrino events detected by neutrino detector 
come from the observed core-collapse events
with certain survey limiting magnitudes.
This panel shows that 
$\sim 18\%$ of the
neutrino events detected above 10 MeV 
are contributed by core-collapse events observed by surveys 
with a $\mag{23}$ limiting magnitude.

We could thus estimate   
the {\em extreme lower limit} to the DSNB
to be 
$\approx 15\%$ of the total detection events
in the 10 -- 18 MeV range,
and $\approx 29\%$  
of the total events in the 18 -- 26 MeV range,
assuming surveys with $\mlim \approx \mag{23}$. 
Surveys including deeper scans will see larger fractions, e.g.,
approaching 
$\approx 54\%$ of the event rate within 18 -- 26 MeV 
for $\mlim \approx \mag{25}$.
Notice that the numbers we showed above might be slightly 
lower than the percentages
read directly from the lower panel of Fig.~\ref{fig:nuobsr}, 
since the numbers above are integrated only through 
the detectable energy range
to reflect the best of what neutrino detectors would observe,
while in Fig.~\ref{fig:nuobsr}, the numbers are 
integrated out to 30 MeV.


\end{document}